\documentclass[twocolumn,preprintnumbers,amsmath,superscriptaddress,amssymb,aps]{revtex4-1}

\usepackage{graphicx}
\usepackage{dcolumn}
\usepackage{bm}
\usepackage{amsmath}
\usepackage{amssymb}
\usepackage{graphicx}
\usepackage{indentfirst}
\usepackage{booktabs}
\usepackage{multirow}
\usepackage{colortbl}

\selectfont

\begin{document}
\title{Multifield tunable valley splitting in two-dimensional MXene Cr$_2$COOH}

\author{Ping Li}
\email{pli@xjtu.edu.cn}
\address{State Key Laboratory for Mechanical Behavior of Materials, Center for Spintronics and Quantum System, School of Materials Science and Engineering, Xi'an Jiaotong University, Xi'an, Shaanxi, 710049, China}
\address{State Key Laboratory for Surface Physics and Department of Physics, Fudan University, Shanghai, 200433, China}
\author{Chao Wu}
\address{State Key Laboratory for Mechanical Behavior of Materials, Center for Spintronics and Quantum System, School of Materials Science and Engineering, Xi'an Jiaotong University, Xi'an, Shaanxi, 710049, China}
\author{Cheng Peng}
\address{State Key Laboratory for Mechanical Behavior of Materials, Center for Spintronics and Quantum System, School of Materials Science and Engineering, Xi'an Jiaotong University, Xi'an, Shaanxi, 710049, China}
\author{Mutian Yang}
\address{State Key Laboratory for Mechanical Behavior of Materials, Center for Spintronics and Quantum System, School of Materials Science and Engineering, Xi'an Jiaotong University, Xi'an, Shaanxi, 710049, China}
\author{Wei Xun}
\email{xunwei@hyit.edu.cn}
\address{Faculty of Electronic Information Engineering, Huaiyin Institute of Technology, Huaian 223003, China}

\date{\today}

\begin{abstract}
Manipulation of the valley degree of freedom provides a novel paradigm in quantum information technology. Here, through first-principles calculations and model analysis, we demonstrate that monolayer Cr$_2$COOH MXene is a promising candidate material for valleytronics applications. We reveal that Cr$_2$COOH is a ferromagnetic semiconductor and harbors valley features. Due to the simultaneous breaking inversion symmetry and time-reversal symmetry, the valleys are polarized spontaneously. Moreover, the valley polarization is sizeable in both the valence and conduction bands, benefiting the observation of the anomalous valley Hall effect. More remarkably, the valley splitting can be effectively tuned by the magnetization direction, strain and ferroelectric substrate. More interestingly, the ferroelectric substrate Sc$_2$CO$_2$ can not only regulate the MAE, but also tune valley polarization state. Our findings offer a practical way for realizing highly tunable valleys by multiferroic couplings.
\end{abstract}

\maketitle
\section{Introduction}
In addition to charge and spin, the valley index is rapidly emerging as a distinctive electronic degree of freedom of electrons, which has simulated intensive investigation recently \cite{1,2,3,4}. The valley generally refers to the local energy extremum point in the valence band or conduction band. The valley index is especially robust in terms of impurity and phonon scatterings due to the large separation in momentum space \cite{5}. In order to make use of the valley index as an information carrier, it is necessary to manipulate the carriers in the valleys, thereby generating valley polarization. The external magnetic field or the proximity-induced Zeeman effect has been proven to be effective in tuning valley degeneracy for transition metal dichalcogenides monolayer \cite{6,7,8,9}. There is an intriguing question, whether a valley in a two-dimensional (2D) material can be modulated by both strain and magnetic means. If so, the valley degree of freedom would be highly adjustable and have huge applications in valleytronics. However, there are few examples of strain and magnetic simultaneous control of the valley, which is worthy of further exploration \cite{10,11}.

Recently, a series of MXenes with the general formula M$_{n+1}$C$_n$X (M = transition metal, X = F, Cl, OH, or O, n = 1-4) have been exfoliated from their MAX (M is an transition metal, A is mainly a group IIIA or IVA element, X is C and/or N) \cite{12} phases to their 2D limit \cite{12,13,14,15,16}. They have various excellent properties, such as intrinsic ferromagnetism \cite{17}, topological insulating \cite{18}, and superconductivity \cite{19}. The interplay between spin and valley in 2D magnetic materials may produce spontaneous valley polarization, without the need for annoying external methods. So far, only a few candidate systems have been proposed, such as VSe$_2$, Cr$_2$Se$_3$, XY (X = K, Rb, Cs; Y = N, P, As, Sb, Bi), MnPSe$_3$, VSiXN$_4$ (X = C, Si, Ge, Sn, Pb) \cite{20,21,22,23,24,25}. Therefore, if the valley degree of freedom is found in magnetic MXene monolayer, it becomes more attractive. It is likely to realize coupling between magnetic order and the valley degree of freedom. It will be further regulated by multiple physical fields.

In this work, using first-principles calculations, we present the discovery of intrinsic spontaneous valley polarization in 2D Cr$_2$COOH MXene. Our results show that monolayer Cr$_2$COOH is a semiconductor with a moderate band gap locating at the K and K' points, forming two inequivalent valleys. When the spin-orbit coupling (SOC) is switched on, the valleys of Cr$_2$COOH show a large valley spin splitting. More interestingly, the large valley splitting occurs not only at the valence band maximum (VBM), but also show at the conduction band minimum (CBM), which is different from that of the typical 2D ferrovalley materials. Subsequently, the multiferroic couplings between ferromagnetism and ferrovalley are found. Importantly, the valley splitting can be tuned by the strain, magnetization rotation and ferroelectric substrate. We also confirm the anomalous valley Hall effect (AVHE) can be realized in 2D Cr$_2$COOH. The highly tunable valley splitting in multiferroic MXene provides a practical avenue for designing advanced spintronic and valleytronic devices based on the couplings between multiferroic orders.

\section{STRUCTURES AND COMPUTATIONAL METHODS}
To explore the electronic and magnetic structures, we used the Vienna $Ab$ $initio$ Simulation Package (VASP) \cite{26,27,28} within the framework of the density functional theory (DFT) for the first-principles calculations. The exchange-correlation energy was described by the generalized gradient approximation (GGA) with the Perdew-Burke-Ernzerhof (PBE) functional \cite{29}. The plane-wave basis with a kinetic energy cutoff of 500 eV was employed, and $21\times 21\times 1$ and $30\times 30\times 1$ $\Gamma$-centered $k$ meshes were adopted for structural optimization and self-consistent calculations. A vacuum of 20 $\rm \AA$ was set along the c-axis, to avoid the interaction between the sheet and its periodic images. The total energy convergence criterion and the force were set to be 10$^{-6}$ eV and -0.01 eV/$\rm \AA$, respectively. To describe strongly correlated 3d electrons of Cr \cite{30}, the GGA + U method is applied. The Coulomb repulsion U is varied between 0 eV and 5 eV. To investigate the dynamical stability, the phonon spectra were calculated using a finite displacement approach as implemented in the PHONOPY code \cite{31}. $\emph{Ab initio}$ molecular dynamic (AIMD) simulation is carried out in a canonical ensemble based on the algorithm of Nose at a temperature of 300 K with a $3\times 3\times 1$ supercell \cite{32,33,34}. The maximally localized Wannier functions (MLWFs) were employed to construct an effective tight-binding Hamiltonian to explore the Berry curvature, anomalous Hall conductivity (AHC), and edge states \cite{35}. The calculated the AHC, it performed the Berry curvature calculations using the formula
\begin{equation}
	\sigma_{xy} = -\frac{e^2}{h}\int_{BZ}\frac{d^3k}{(2\pi)^3}\Omega^z(\textbf{k}),
\end{equation}

\begin{equation}
	\Omega^z(\textbf{k})=-\sum_{n}f_{n}\sum_{n\prime \neq n}\frac{2Im \left \langle \psi_{nk} \mid v_{x} \mid \psi_{n\prime k} \right \rangle \left \langle \psi_{n\prime k} \mid v_{y} \mid \psi_{nk} \right \rangle}{(E_{n\prime}-E_{n})^2},
\end{equation}
where $\Omega^z(\textbf{k})$ is the Berry curvature in the reciprocal space, $v_{x}$ and $v_{y}$ are operator components along the x and y directions, and $f_{n}=1$ for the occupied bands, respectively \cite{36,37,38}.

\section{RESULTS AND DISCUSSION }	
\subsection{Structure and stability}
As shown in Figure 1(a), it shows the crystal structure of monolayer Cr$_2$COOH MXene. The monolayer has 2D hexagonal lattice, and consists of six atomic layers stacked as H-O-Cr-C-Cr-O. Therefore, Cr$_2$COOH exhibits the intrinsic broken inversion symmetry. The in-plane lattice constant of its is optimized to 2.99 $\rm \AA$ and the thickness of the monolayer is 5.46 $\rm \AA$. To investigate the bonding characteristics of Cr$_2$COOH MXene, we calculate its electron localization function (ELF) in the plane containing H-O-Cr-C-Cr-O \cite{39}. The ELF value is between 0 and 1. 1 means that the electron is completely localized, while 0 indicates that the electron is completely delocalized. As shown in Figure 1(c), the electrons are mainly localized around the atoms, while those in between the atoms are negligible, indicating an ionic bonding for all the bonds. Moreover, the thermal dynamic stability of monolayer Cr$_2$COOH MXene is estimated by the phonon dispersion spectrum and AIMD calculations. As shown in Figure 1(d), The absence of imaginary modes along the high-symmetry lines verifies the dynamical stability of monolayer Cr$_2$COOH MXene. The total energies of monolayer Cr$_2$COOH MXene during 5 ps with a time step of 1 fs at 300 K by AIMD simulations are shown in Figure S1. The small fluctuations of energy and integrity of the original configuration with time evolution demonstrate good thermal stability of monolayer Cr$_2$COOH MXene.

\begin{figure}[htb]
\begin{center}
\includegraphics[angle=0,width=1.0\linewidth]{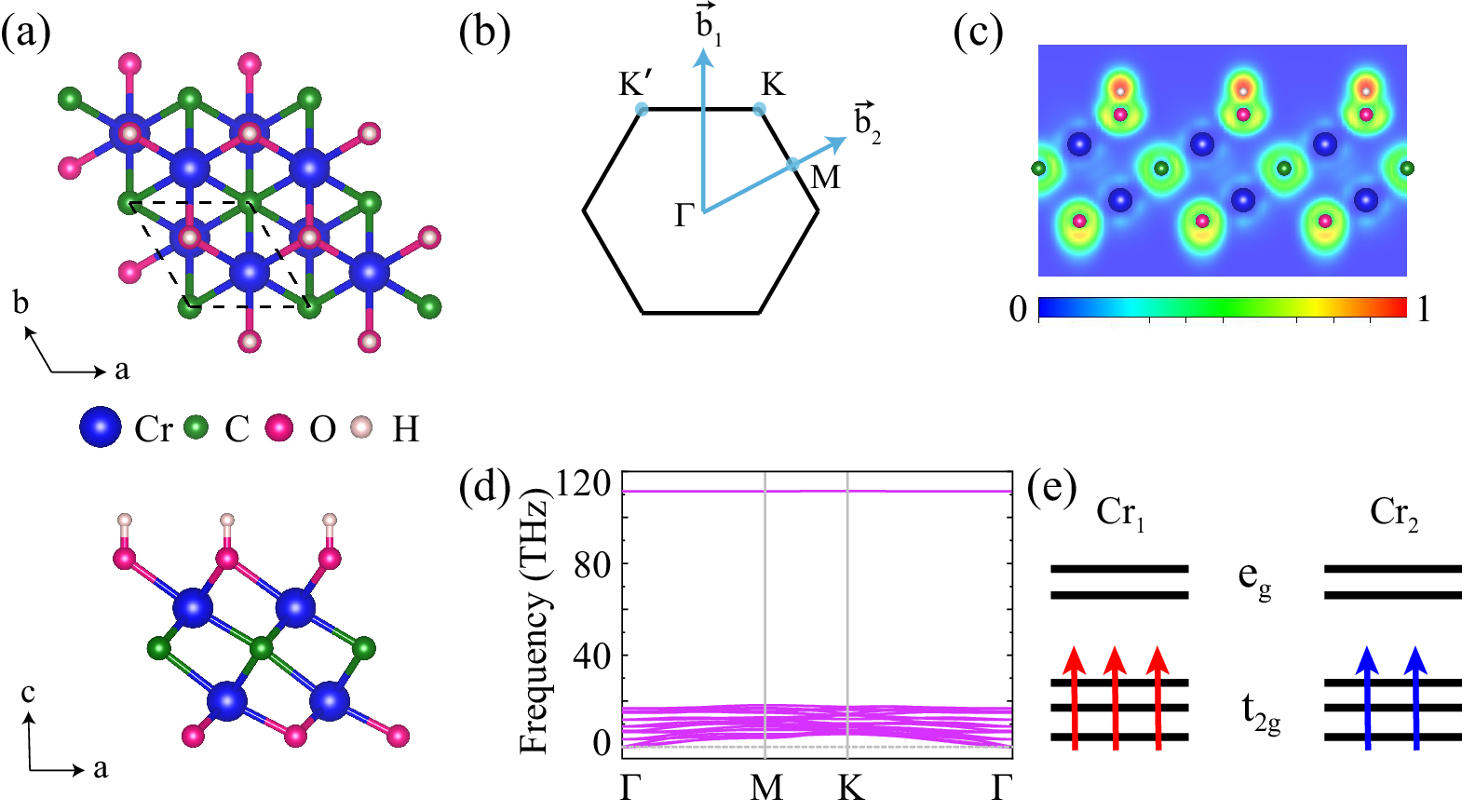}
\caption{(a) The top and side views of the crystal structure for Cr$_2$COOH monolayer. The blue, green, magenta, and white balls represent Cr, C, O, and H elements, respectively. (b) The Brillouin zone (BZ) of the honeycomb lattice with the reciprocal lattice vectors $\vec{b}_1$ and $\vec{b}_2$. $\Gamma$, K, and M are high-symmetry points in the BZ. (c) Electron localization function of Cr$_2$COOH monolayer. (d) Splitting of $\emph{d}$ orbitals of the Cr atom under the distorted octahedral crystal field. (e) The calculated phonon dispersion curves along the high-symmetry lines.
}
\end{center}
\end{figure}

\subsection{Magnetic property}
The valence electronic configuration of the Cr atom is 3$\emph{d}$$^5$4$\emph{s}$$^1$. For the upper Cr atom (Cr$_1$), it would lose three electrons to the six neighboring C and O atoms because of its surroundings, leading to the electronic configuration of 3$\emph{d}$$^3$4$\emph{s}$$^0$. While for the lower Cr atom (Cr$_2$), due to the different coordinated atoms, one more electron would be transferred to the neighboring atoms, and the electronic configuration of Cr$_2$ would be 3$\emph{d}$$^2$4$\emph{s}$$^0$. According to Hund's rule and the Pauli exclusion principle, the electron configuration of Cr$_1$ would half fill the t$_2g$ orbitals, while the Cr$_2$ would half-fill two of these three orbitals, as shown in Figure 1(e). The Cr$_2$COOH MXene has a total magnetic moment of 5 $\mu_B$ per unit cell, while Cr$_1$ and Cr$_2$ yield magnetic moment of 3 $\mu_B$ and 2 $\mu_B$, respectively.

\begin{figure}[htb]
\begin{center}
\includegraphics[angle=0,width=1.0\linewidth]{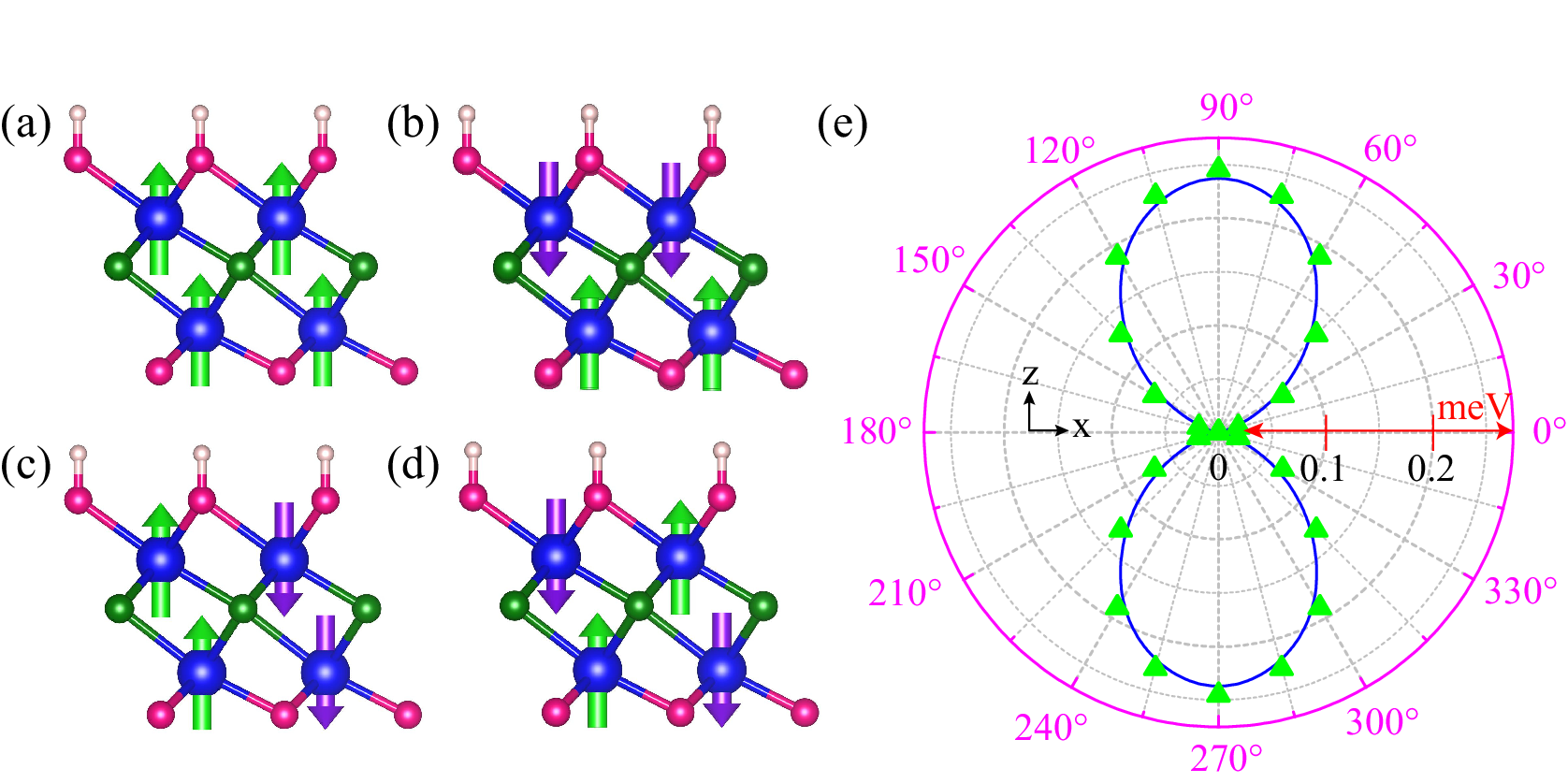}
\caption{(a) FM, (b) AFM1, (c) AFM2, and (d) AFM3 configurations in Cr$_2$COOH monolayer. (e) Angular dependence of the MAE of Cr$_2$COOH with the direction of magnetization lying on the xz plane.
}
\end{center}
\end{figure}

To determine the magnetic ground state of the monolayer Cr$_2$COOH MXene, four possible magnetic configurations are considered, namely, the ferromagnetic (FM) and three types of antiferromagnetic ones (AFM1, AFM2, AFM3) using the GGA + U method [see Figure. 2(a-d)]. To determine a reasonable value of U, the Coulomb repulsion U is varied between 0 eV and 5 eV. As shown in Figure S2(a), The FM state is the ground state in all case except U = 0 eV. It is well known that transition metal atoms must be modified by Hubbard U in the DFT. In previous reports, the Hubbard U is often taken to be 3 eV for the Cr atom \cite{40,41}. Therefore, we choose U$_{eff}$ = 3 eV to investigate the monolayer Cr$_2$COOH MXene. The FM state is 362.5 meV, 1196.6 meV, and 1315.5meV lower in energy than the AFM1, AFM2, and AFM3 states, respectively. The FM ground state of the monolayer Cr$_2$COOH MXene can be understood by studying the crystal structure. In Cr$_2$COOH, the Cr-C-Cr bond angles are 88.4$^\circ$, which is close to 90.0$^\circ$. According to the Goodenough-Kanamori-Anderson rule \cite{42,43,44}, this configuration is beneficial to FM coupling.

The magnetic anisotropy energy (MAE) is very important to determine the thermal stability of magnetic ordering. The MAE mainly originates from the SOC interaction \cite{45,46}. Heavy elements are more preferred because their strong SOC effect can give rise to a large MAE. It is defined as MAE = E$_{100}$ - E$_{001}$, where E$_{100}$ and E$_{001}$ represents the total energy of the magnetic moment along [100] and [001], respectively. The negative value of the MAE shows the easy axis is along the x axis rather than along the z axis. The MAE is -0.245 meV per unit cell, indicating the magnetization along the x axis. With such a small MAE, the direction of magnetization can be easily controlled out-of-plane by a small external magnetic field. For hexagonal lattice magnetization direction along (100), the magnetic field (H = -$\frac{4}{3}\frac{K}{\mu_0M}$)is required to tune the direction of in-plane magnetization to out-of plane magnetization \cite{47}. Where K is the MAE, $\mu_0$ is permeability of vacuum, and M is the total magnetic moment of the system. Thus, adjusting the magnetization direction of Cr$_2$COOH out-of-plane requires 1.13 T. Importantly, we found that Hubbard U had almost no effect on the MAE, as shown in Figure S2(b).

Based on the octahedral symmetry of monolayer Cr$_2$COOH MXene, the angular dependence of the MAE can be described using the equation:
\begin{equation}
\rm MAE= K_1 cos^2\theta + K_2 cos^4\theta,
\end{equation}
where K$_1$ and K$_2$ are the anisotropy constants and $\theta$ is the azimuthal angle of rotation. If K$_1$ $>$ 0, the preferred magnetization direction will be along an in-plane easy axis (x-axis), while K$_1$ $<$ 0 suggests that it will be perpendicular to the z-axis. Our calculated MAEs display a good fit of Eq (3) as presented in Figure 2(e), it clearly shows that the MAE strongly depends on the direction of magnetization in the xz plane.

In addition, we calculated the Curie temperature of Cr$_2$COOH by the Monte Carlo simulations, as shown in Figure S3. The Monte Carlo simulations were implemented on a 60 $\times$ 60 lattice grid lasting for 100000 steps with a step size of 1 K for the calculation of Curie temperature. Although the U value has an effect on the Curie temperature, the Curie temperature is above 700 K. This is perfectly suited to the application of spintronic devices.

\subsection{Electronic band structure}
\begin{figure*}[htb]
\begin{center}
\includegraphics[angle=0,width=0.8\linewidth]{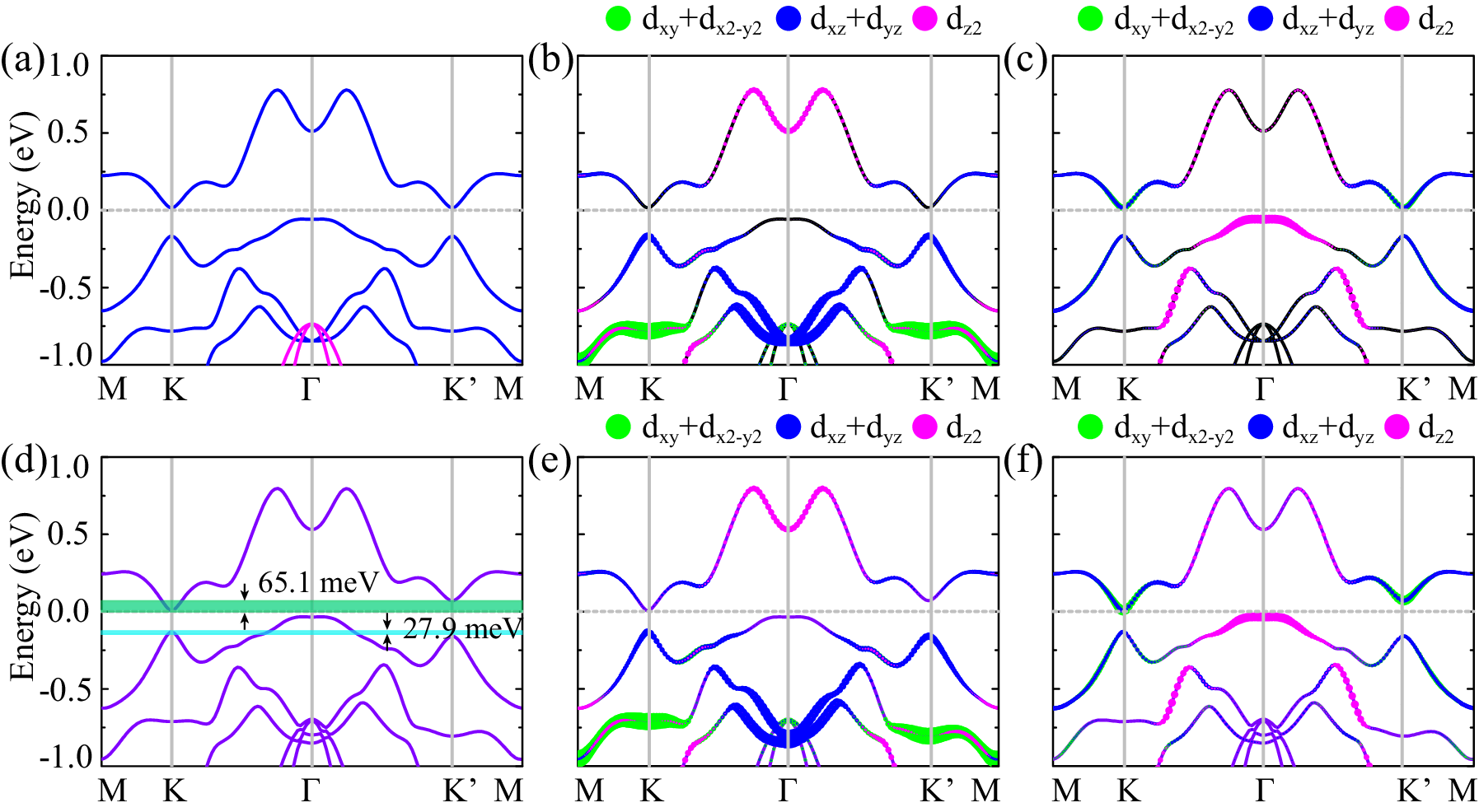}
\caption{ Band structures of Cr$_2$COOH monolayer (a) without and (d) with considering the SOC. Band structures of Cr$_2$COOH monolayer without considering SOC projected on the $\emph{d}$ orbitals of (b) Cr$_1$ and (c) Cr$_2$ atoms and with considering SOC projected on the $\emph{d}$ orbitals of (e) Cr$_1$ and (f) Cr$_2$ atoms. The blue and magenta lines represent spin up and spin down bands, respectively. The valence and conduction band valley splitting are indicated by the light blue and green shading, respectively.
}
\end{center}
\end{figure*}

In the following, we focus on band structures and associated valley properties of monolayer Cr$_2$COOH MXene, which are shown in Figure 3. In the absence of SOC, the band structure around the Fermi level are only from the spin up channel. It is a semiconductor when Hubbard U is less than 3 eV [see Figure S4(a)]. As the U value continues to increase, monolayer Cr$_2$COOH MXene becomes a half-metal. From the orbital-resolved band structure, as shown in Figure 3(b, c), the VBM bands are mainly contributed by Cr$_1$ d$_{xz}$/d$_{yz}$ orbitals, while the CBM bands are dominated by d$_{xy}$/d$_{x2-y2}$ of the Cr$_2$ atom. In order to show more clearly the orbital composition of the VBM and CBM of K point, we calculated the charge density. As shown in Figure S5, this is almost identical to the orbital-resolved band structure. When the SOC is included, as shown in Figure 3(d), the degeneracy between K and K' valleys is lifted. In detail, the K valley shifts above the K' valley in the valence band, while the K valley shifts below the K' valley in the conduction band. Hence, the valley polarization is realized spontaneously in both the valence and conduction bands of monolayer Cr$_2$COOH MXene. The total valley splitting is 93 meV in the Cr$_2$COOH monolayer, which is larger than that in other ferrovalley materials, e. g. VSe$_2$ (90 meV) \cite{20}, Cr$_2$Se$_3$ (18.7 meV) \cite{21}, VSiXN$_4$ ($\thicksim$ 70 meV) \cite{24}, and YCl$_2$ (22.35 meV) \cite{48}. This huge valley splitting is promising to be observed experimentally. The spontaneous valley polarization originates from the combined effect of magnetic exchange interaction and SOC. Importantly, we found that the spontaneous valley polarization increases gradually with the increase of U value [see Figure S2(c, d)]. The change of band gap with U value is listed in Table SI.

Interestingly, the spontaneous valley polarization of monolayer Cr$_2$COOH MXene is different from most of the previously investigated 2D valleytronic materials \cite{20,21,24,48,49}, wherein a sizeable spontaneous valley polarization usually occurs in either the valence band (VSe$_2$ \cite{20}, VSiXN$_4$ \cite{24}, YX$_2$ \cite{48}) and GdI$_2$ \cite{49}, or conduction band (Cr$_2$Se$_3$ \cite{21}), but not both.

In order to understand the underlying mechanism for the ferrovalley effect in Cr$_2$COOH, we tried to construct an effective Hamiltonian model based on the DFT calculations. we take the SOC effect as the perturbation term, which is
\begin{equation}
\hat{H}_{SOC} = \lambda \hat{S} \cdot \hat{L} = \hat{H}_{SOC}^{0} + \hat{H}_{SOC}^{1},
\end{equation}
where $\hat{S}$ and $\hat{L}$ are spin angular and orbital angular operators, respectively. $\hat{H}_{SOC}^{0}$ and $\hat{H}_{SOC}^{1}$ represent the interaction between the same spin states and between opposite spin states, respectively. Since the VBM and CBM belong to the spin up bands, for monolayer Cr$_2$COOH MXene, the single valley is composed of only one spin channel [see Figure 3(a)], and the other spin channel is far from the valleys. Hence, the term $\hat{H}_{SOC}^{1}$ can be ignored. On the other hand, $\hat{H}_{SOC}^{0}$ can be written in polar angles
\begin{equation}
\hat{H}_{SOC}^{0} = \lambda \hat{S}_{z'}(\hat{L}_zcos\theta + \frac{1}{2}\hat{L}_+e^{-i\phi}sin\theta + \frac{1}{2}\hat{L}_-e^{+i\phi}sin\theta),
\end{equation}
In the out-of-plane magnetization case, $\theta$ = $\phi$ = 0$^ \circ$, then the $\hat{H}_{SOC}^{0}$ term can be simplified as
\begin{equation}
\hat{H}_{SOC}^{0} = \lambda \hat{S}_{z} \hat{L}_z,
\end{equation}

Considering the orbital contribution around the valleys and the C$_3$ symmetry, we adopted $|$$\psi$$_v$$^{\tau}$$\rangle$=$\frac{1}{\sqrt{2}}$($|$d$_{xz}$$\rangle$+i$\tau$$|$d$_{yz}$$\rangle$)$\otimes$$|$$\uparrow$$\rangle$, $|$$\psi$$_c$$^{\tau}$$\rangle$=$\frac{1}{\sqrt{2}}$($|$d$_{xy}$$\rangle$+i$\tau$$|$d$_{x2-y2}$$\rangle$)$\otimes$$|$$\uparrow$$\rangle$ as the orbital basis for the VBM and CBM, where $\tau$ = $\pm$1 indicate the valley index corresponding to $\rm K/\rm K'$. The energy levels of the valleys for the VBM and CBM can be expressed as E$_v$$^ \tau$ = $\langle$ $\psi$$_v$$^ \tau$ $|$ $\hat{H}$$_{SOC}^{0}$ $|$ $\psi$$_v$$^ \tau$ $\rangle$ and E$_c$$^ \tau$ = $\langle$ $\psi$$_c$$^ \tau$ $|$ $\hat{H}$$_{SOC}^{0}$ $|$ $\psi$$_c$$^ \tau$ $\rangle$, respectively. Then, the valley polarization in the valence and conduction bands can be expressed as
\begin{equation}
E_{v}^{K} - E_{v}^{K'} = i \langle d_{xz} | \hat{H}_{SOC}^{0} | d_{yz} \rangle - i \langle d_{yz} | \hat{H}_{SOC}^{0} | d_{xz} \rangle \approx 2\alpha,
\end{equation}
\begin{equation}
E_{c}^{K} - E_{c}^{K'} = i \langle d_{xy} | \hat{H}_{SOC}^{0} | d_{x2-y2} \rangle - i \langle d_{x2-y2} | \hat{H}_{SOC}^{0} | d_{xy} \rangle \approx 4\beta,
\end{equation}
where the $\hat{L}_z|d_{xz} \rangle$ = i$\hbar$$|d_{yz} \rangle$, $\hat{L}_z|d_{yz} \rangle$ = -i$\hbar$$|d_{xz} \rangle$, $\hat{L}_z|d_{xy} \rangle$ = -2i$\hbar$$|d_{x2-y2} \rangle$, $\hat{L}_z|d_{x2-y2} \rangle$ = 2i$\hbar$$|d_{xy} \rangle$, $\alpha = \lambda \langle d_{xz} |\hat{S}_{z'}| d_{xz} \rangle$, and $\beta = \lambda \langle d_{x2-y2} |\hat{S}_{z'}| d_{x2-y2} \rangle$. The analytical result certificates that the valley degeneracy splits for the valence and conduction bands are consistent with our DFT calculations ($E_{v}^{K}$ - $E_{v}^{K'}$ = 27.9 meV, $E_{c}^{K}$ - $E_{c}^{K'}$ = 65.1 meV).

\subsection{Anomalous valley Hall effect}
To characterize the valley-contrasting physics in monolayer Cr$_2$COOH MXene, we calculate its Berry curvature. The Berry curvature $\Omega(k)$ of monolayer Cr$_2$COOH MXene in the along high-symmetry points and entire 2D Brillouin zone is shown in Figure 4(a, b). Clearly, the Berry curvature of monolayer Cr$_2$COOH MXene is peaked around K and K' valleys. The absolute values of Berry curvature at K and K' valleys are different, and the signs are opposite. It indicated the simultaneous breaking of inversion symmetry and time-reversal symmetry in monolayer Cr$_2$COOH MXene. By integrating the Berry curvature over the Brillouin zone, one can further calculate the AHC. As shown in Figure 4(c), when the Fermi level lies between the energies of the valence and conduction bands at the K and K' valleys, as denoted the light blue and green shadows, a fully spin and valley polarized Hall conductivity is generated.

\begin{figure}[htb]
\begin{center}
\includegraphics[angle=0,width=1.0\linewidth]{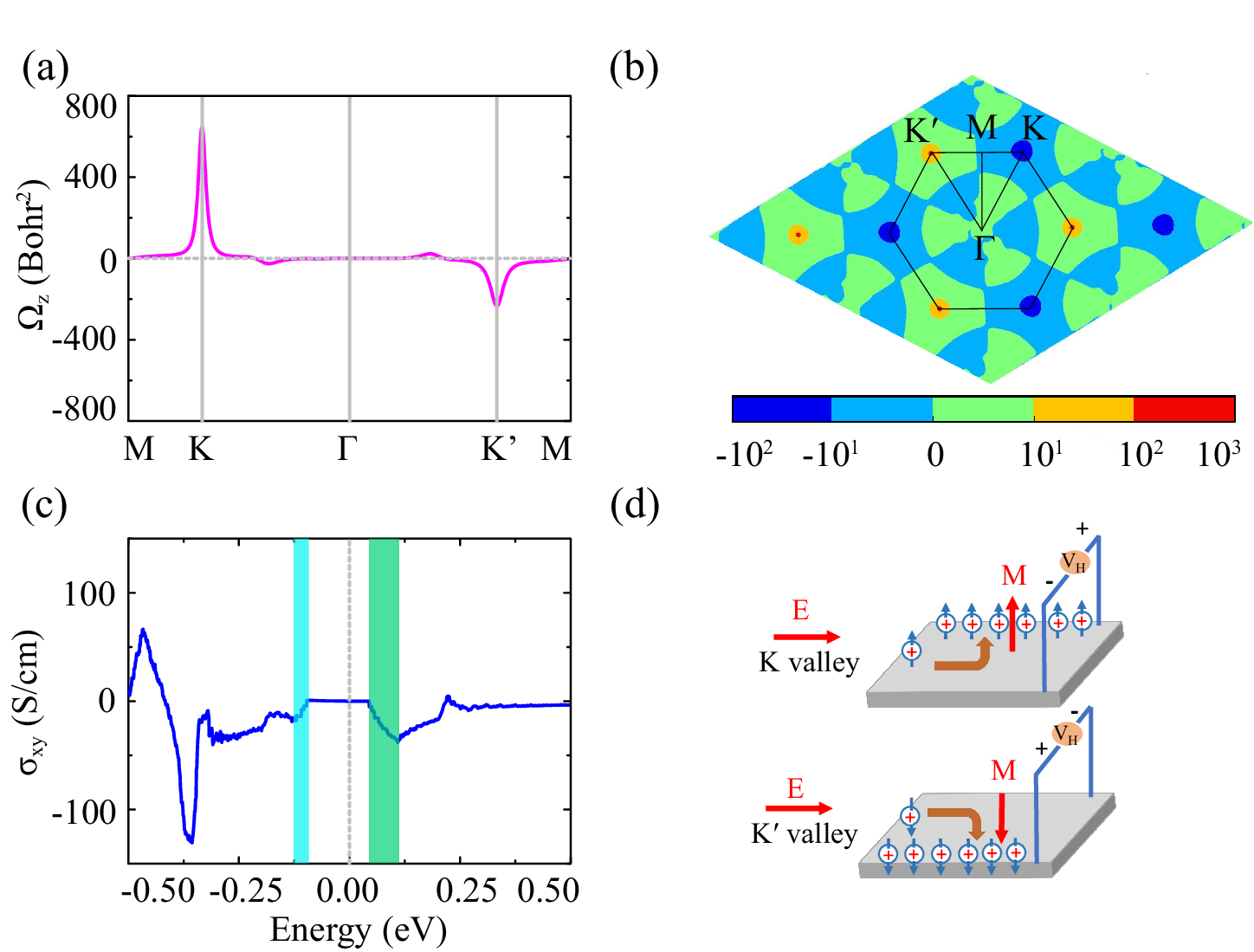}
\caption{(a) The Berry Curvatures of Cr$_2$COOH along the high-symmetry line and in the Brillouin zone (b). (c) Calculated AHC $\sigma_{xy}$ as a function of Fermi energy. The light blue and green shadows denote the valence and conduction band valley splitting between the K and K' valley. (d) Schematic diagram of tunable the anomalous valley Hall effect in hole-doped Cr$_2$COOH monolayer at the K and $\rm K'$ valley, respectively. The holes are denoted by the + symbol. Upward arrows and downward arrows refer to the spin up and spin down carriers, respectively.
}
\end{center}
\end{figure}

The AVHE of monolayer Cr$_2$COOH MXene is illustrated in Figure 4(d). With the hole-doping condition, when the magnetism direction of monolayer Cr$_2$COOH MXene is in the +z direction, the spin up holes from the $\rm K'$ valley will be generated and accumulate on one boundary of the sample under an in-plane electrical field [upper plane of Figure 4(d)]. On the other hand, when the magnetism direction is in the -z direction, the spin-up holes from the K valley will be generated and accumulate on the opposite boundary of the sample under an in-plane electrical field [lower plane of Figure 4(d)]. This feature shows that monolayer Cr$_2$COOH MXene is an ideal candidate for the high-performance valleytronics devices.

\subsection{Magnetic tune valley splitting}
Considering that the monolayer Cr$_2$COOH MXene is found to be a ferrovalley material and it has ferromagnetic orders as well, we will further explore the coupling between these multiferroic orders. Firstly, we investigate the effect of the magnetization on the valley splittings. As shown in Figure 5, it exhibits the valley splitting of the valence and conduction bands as a function of the magnetization direction. With rotating the magnetization from the in-plane direction (0$^ \circ$) to the +z direction (90$^ \circ$), the valley splitting of valence and conduction bands gradually increase from 0 meV to 27.9 meV, and 65.1 meV, respectively. As we continue rotating the magnetization to the -x direction (180$^ \circ$), the magnitude of the valley splitting gradually decreases from the maximum to zero. Interestingly, the variation trend of valley splitting with angle is consistent that of the angular dependence of the MAE. Moreover, the valley splitting of valence and conduction bands is entirely agreement with the two-fold relationship. It further proves the validity of our established effective Hamiltonian for valley splitting.

\begin{figure}[htb]
\begin{center}
\includegraphics[angle=0,width=0.8\linewidth]{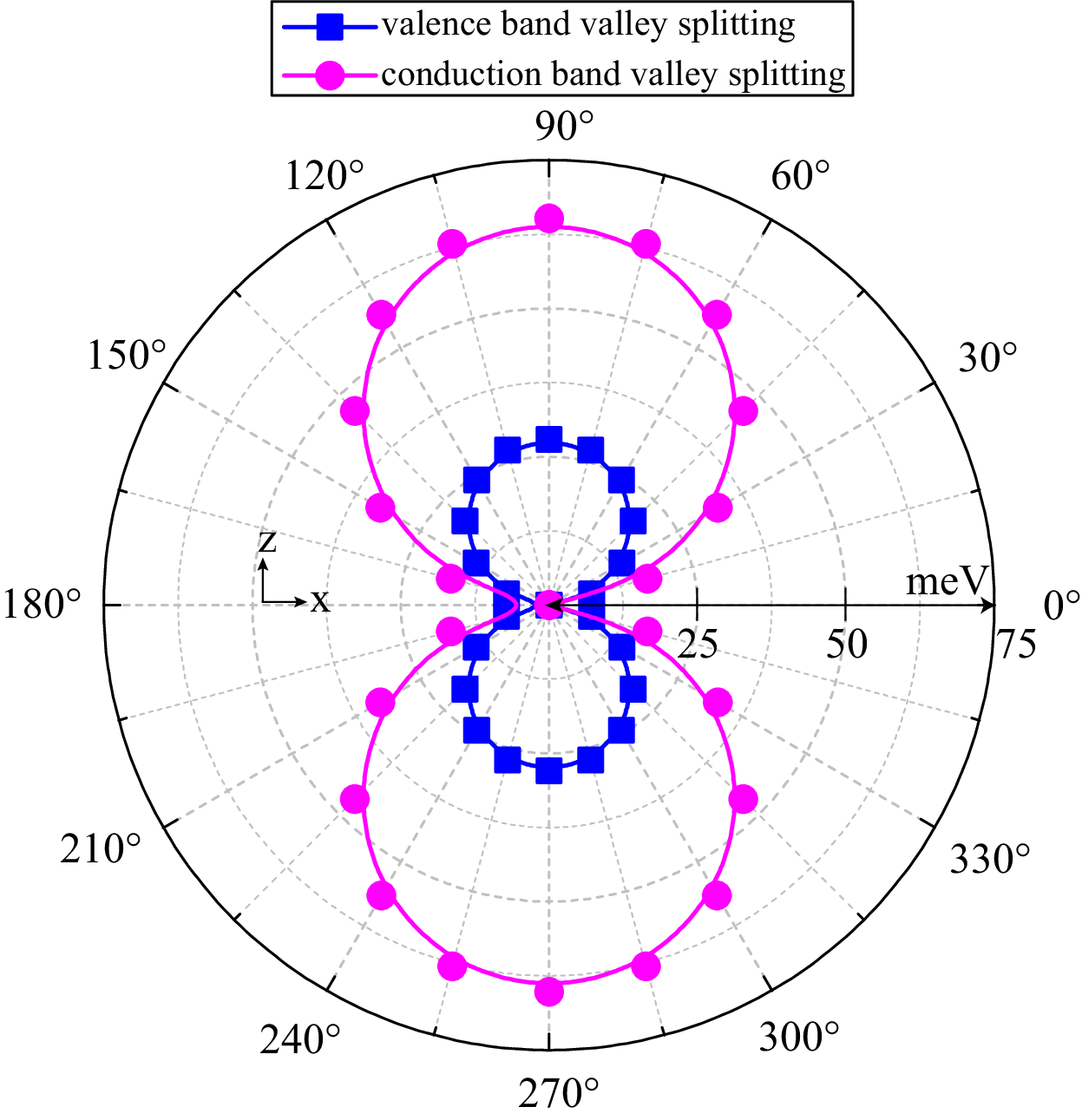}
\caption{ Angular dependence of the valence and conduction band valley splitting of Cr$_2$COOH monolayer with the direction of magnetization lying on the xz plane is indicated by magenta and blue line, respectively.
}
\end{center}
\end{figure}

\subsection{Strain tune valley splitting}
\begin{figure}[htb]
\begin{center}
\includegraphics[angle=0,width=1.0\linewidth]{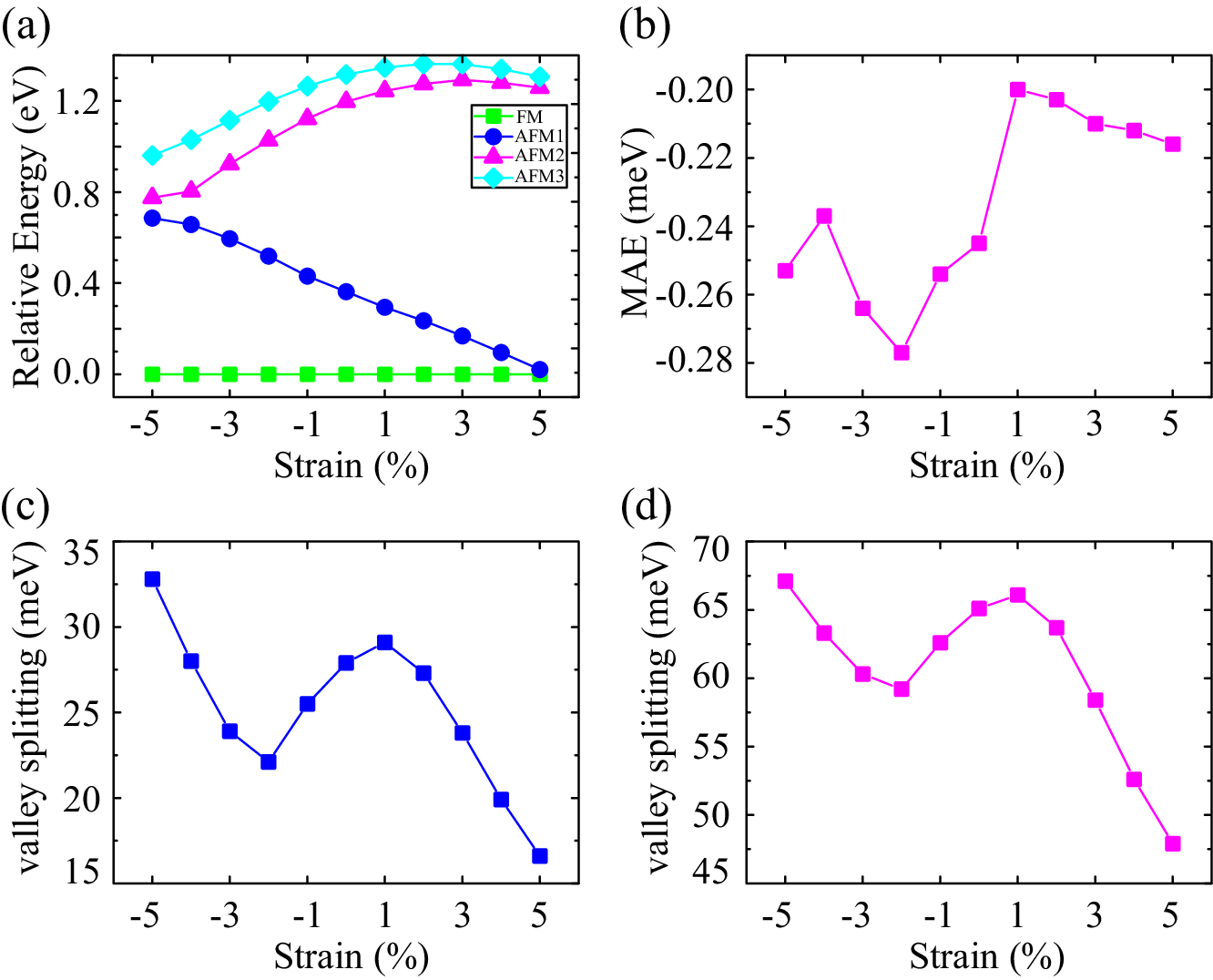}
\caption{ (a) The total energies of Cr$_2$COOH different magnetic structures as a function strain, which are defined relative to that of the FM state. (b) The magnetic anisotropy energy as a function of strain. (c) The valence band valley splitting as a function of strain. (d) The conduction band valley splitting as a function of strain.
}
\end{center}
\end{figure}

In device fabrication, a 2D material is often supported by a substrate which is likely to introduce strain to a 2D material due to lattice mismatch. T. Liu $\emph{et al.}$ observed that substrate-induced strain can modulate the spin dynamics in 2D transition metal dichalcogenide \cite{50}. In addition, we also found that substrate-induced strain can tune spin Hall angle in 2D Bi$_2$Se$_3$ \cite{51}. In the following, we investigate the effect of biaxial strain on the magnetic ground state, valley splitting, and MAE of monolayer Cr$_2$COOH MXene. In the calculations, the biaxial strain is defined as $\varepsilon$ = (a-a$_0$)/a$_0$$\times$100$\%$. In the formula, a and a$_0$ represent lattice constant after and before in-plane biaxial strain is applied, respectively. Calculations are performed for the biaxial in-plane strain in the range of -5$\%$ $\thicksim$ 5$\%$, in which the positive and negative refer to the tensile and compressive strain, respectively. As shown in Figure 6(a), it indicates that the FM ground state is very robust. Moreover, the MAE varies within a range -0.28 meV to -0.20 meV in the strain [see Figure 6(b)]. The valley splitting went through three stages in the strain [see Figure 6(c, d)]. When $\varepsilon$ $<$ -2 $\%$, the valley splitting of valence and conduction bands first decrease. Continuing applying the strain in -2$\%$ $\thicksim$ 1$\%$, the valley splitting then increase. When the strain becomes $\varepsilon$ $>$ 1 $\%$ tensile strain, the valley splitting reduces again. In addition, the band structures vary with strain for monolayer Cr$_2$COOH MXene is summarized in Figure S6.

\subsection{Ferroelectric substrate tune valley splitting}

In order to study valley polarization under the direction of magnetization along the out-of-plane, we investigate that the ferroelectric substrate Sc$_2$CO$_2$ tune the MAE of monolayer Cr$_2$COOH. We noticed that the $\sqrt{3}$ $\times$ $\sqrt{3}$ unit cell of Cr$_2$COOH is commensurate to the 2 $\times$ 2 Sc$_2$CO$_2$ with a lattice mismatch of less than 1$\%$. We have considered four typical configurations, i.e., the H atom of Cr2COOH being in the bridge, hollow, top-O, and top-Sc positions of the Sc2CO2, respectively, as shown in Figure 7(a) and Figure S7(a-c). As listed in Table SII, the total energy calculations show that the top-O and hollow configurations is the most stable for polarization down and polarization up, respectively. We will focus on these two structures in the following discussion. We found a novel phenomenon. When the monolayer Sc$_2$CO$_2$ is polarized upward, the spin-polarized band structures of all structures exhibit semiconducting properties [see Figure 7(b), Figure S7(d-f)]. However, for the polarization downward configuration, the band gap is closed under polarization field, as shown in Figure 7(d), and Figure S7(g-i). More importantly, the ferroelectric substrate Sc2CO2 tune the direction of easy magnetization of Cr2COOH from in-plane to out-of-plane (see Table SIII). We also found that the degeneracy between K and K' valley disappear for polarization upward configuration, when the SOC is included. As shown in Figure 7(c) and Figure S8(a-c), the K' valley shifts above the K valley in the valence band, while the K' valley shifts below the K valley in the conduction band. This is the exact opposite of freestanding Cr$_2$COOH.

\begin{figure}[htb]
\begin{center}
\includegraphics[angle=0,width=1.0\linewidth]{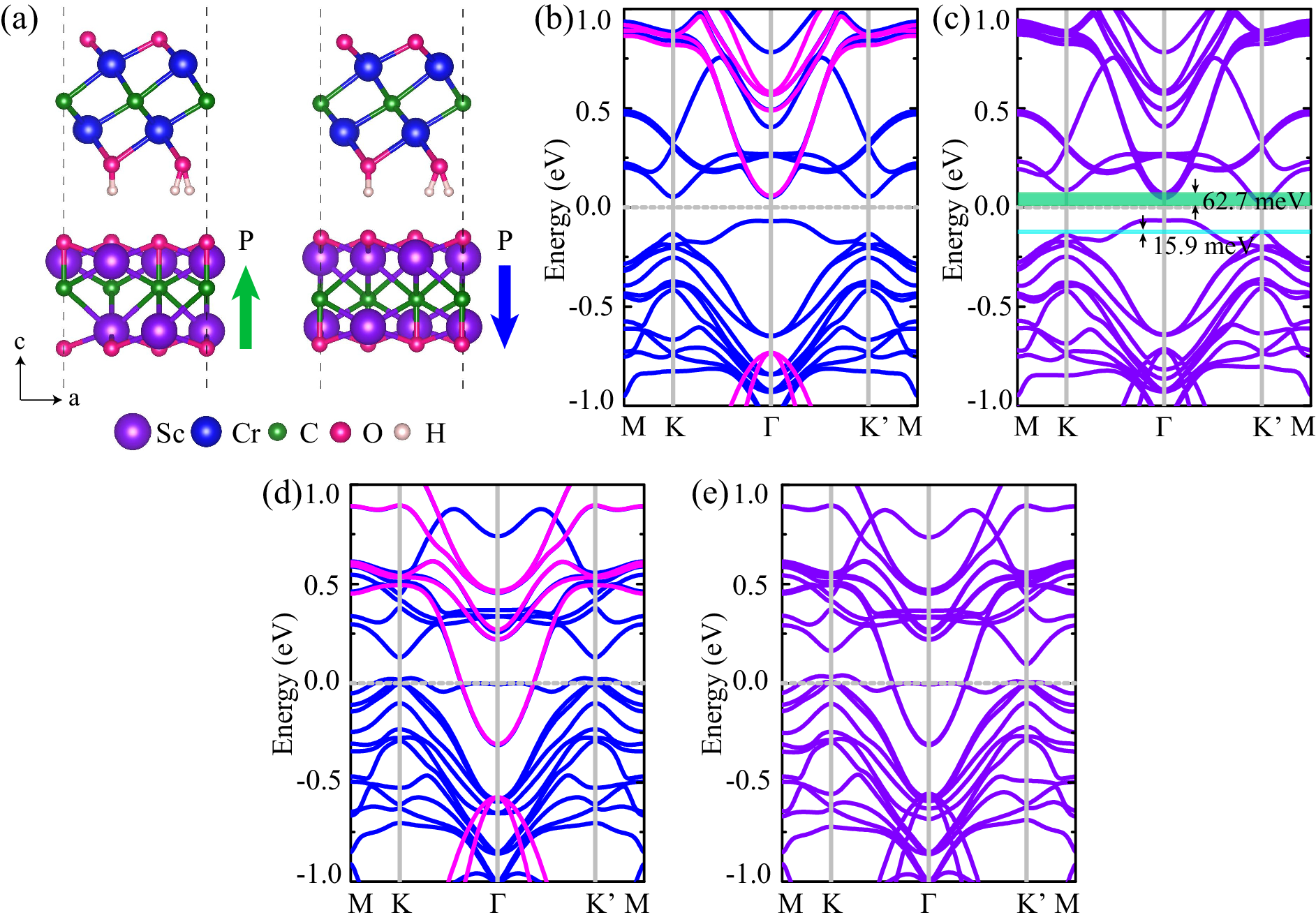}
\caption{ (a) Side views of Cr$_2$COOH/Sc$_2$CO$_2$ heterostructures with the top-O configuration. (b, d) Spin-polarized band structure of Cr$_2$COOH/Sc$_2$CO$_2$ heterostructures for (b) polarization up and (d) polarization down with the top-O configuration, respectively. The blue and magenta lines represent spin up and spin down bands, respectively. (c, e) Band structure with SOC of Cr$_2$COOH/Sc$_2$CO$_2$ heterostructures for (c) polarization up and (e) polarization down with the top-O configuration, respectively.
}
\end{center}
\end{figure}

\section{CONCLUSION}
In conclusion, we propose a novel 2D valleytronic material in monolayer Cr$_2$COOH MXene based on first-principles calculation. We find that monolayer Cr$_2$COOH MXene is a FM semiconductor with the abundant valley physics. In contrast to 2D valleytronic systems reported in previous works, the valley polarization is sizeable in both the valence and conduction bands. This sizeable valley polarization is in favor of realizing the AVHE in monolayer Cr$_2$COOH MXene. Moreover, it is found that the valley splitting can be modulated by the magnetization direction, strain and ferroelectric substrate. When the monolayer Sc$_2$CO$_2$ is polarized upward, the system exhibit valley polarized semiconductor. However, for the polarization downward configuration, the band gap is closed under polarization field. Our work provides not only a unique 2D valleytronic materials but also efficient avenues to tune valley polarization.

\section*{ACKNOWLEDGEMENTS}
This work is supported by the National Natural Science Foundation of China (Grant No. 12004295). P. Li thanks China's Postdoctoral Science Foundation funded project (Grant No. 2022M722547), and the Open Project of State Key Laboratory of Surface Physics (No. KF2022$\_$09).


\end{document}